\documentclass[%
 reprint,%
 amssymb, amsmath,%
 aip,cha
]{revtex4-1}

\usepackage{bm}%
\usepackage[colorlinks=true,linkcolor=blue]{hyperref}%

\usepackage{graphicx}
\usepackage{amsmath}
\usepackage{epstopdf}
\usepackage{color}
\usepackage{float}

\newcommand{\pvec}[1]{\vec{#1}\mkern2mu\vphantom{#1}}

\expandafter\ifx\csname package@font\endcsname\relax\else
 \expandafter\expandafter
 \expandafter\usepackage
 \expandafter\expandafter
 \expandafter{\csname package@font\endcsname}%
\fi
\begin{document}

\title{Cluster self-assembly condition for arbitrary interaction potentials}%

\author{Alejandro Mendoza-Coto}%
\affiliation{Departamento de F\'\i sica, Universidade Federal de Santa Catarina, 88040-900 Florian\'opolis, Brazil\looseness=-1}%

\author{R\'omulo Cenci}%
\affiliation{Departamento de F\'\i sica, Universidade Federal de Santa Catarina, 88040-900 Florian\'opolis, Brazil\looseness=-1}%

\author{Guido Pupillo}%
\affiliation{icFRC, ISIS (UMR 7006), IPCMS (UMR 7504), Universit\'e de Strasbourg and CNRS, 67000 Strasbourg, France\looseness=-1}%

\author{Rogelio D\'iaz-M\'endez}%
\affiliation{Department of Physics, KTH Royal Institute of Technology, SE-10691 Stockholm, Sweden\looseness=-1}%

\author{Egor Babaev}%
\affiliation{Department of Physics, KTH Royal Institute of Technology, SE-10691 Stockholm, Sweden\looseness=-1}%


\begin{abstract}
	We present a sufficient criterion for the emergence of cluster phases in an ensemble of interacting classical particles with repulsive two-body interactions.     
	Through a zero-temperature analysis in the low density region we determine the relevant characteristics of the interaction potential that make the energy of a two-particle cluster-crystal become smaller than that of a simple triangular lattice in two dimensions.
	The method leads to a mathematical condition for the emergence of cluster crystals in terms of the sum of Fourier components of a regularized interaction potential, which can be in principle applied to any arbitrary shape of interactions.
	We apply the formalism to several examples of bounded and unbounded potentials with and without cluster-forming ability.
	In all cases, the emergence of self-assembled cluster crystals  is well captured by the presented analytic criterion and verified with known results from molecular dynamics simulations at vanishingly temperatures. Our work generalises known results for bounded potentials to repulsive potentials of arbitrary shape.
\end{abstract}

\maketitle


\section{Introduction}

The self-assembly of interacting particles into cluster phases has been the subject of intense research in recent years, as different communities have focused on the detailed study of pattern formation in many-particle systems.
In most cases, from colloids and polymers~\cite{zhuang17,lindquist17} to  vortex matter in type-1.5 superconductors,~\cite{bs11,Babaev.Carlstrom.ea:10,silaev1,moshchalkov,Ray.Gibbs.ea:14} cold atoms~\cite{cinti14,dm15,pupillo20} and nuclear pasta,~\cite{caplan18} the emergence of cluster phases is understood as an expression of the complex energy landscape generated by a large class of pair-wise potentials commonly referred to as cluster-forming interactions (CFI). 
Seminal works~\cite{likos01,likos01b} for {\it bounded} interactions have rigorously shown that clusters of overlapping particles can be formed in a fluid at sufficiently high densities and can crystallize upon further compression into lattices with density-independent lattice constants. 
These cluster crystals can appear as equilibrium configurations in two  and three dimensions~\cite{zhang10,dm19w1pre} with a number of particles per cluster that increases with increasing density.~\cite{likos07,likos08, glaser07}
The requirement for these bounded potentials to be cluster-forming is that their Fourier transform becomes negative at some wave vector $k$, at which the system can crystallize. 

However, no such criterion exists in general for unbounded interactions or for the regime of intermediate densities of a system of bounded interactions whose Fourier transform does not display any negative component.~\cite{likos01,caprini18} 
Furthermore, results have not been generally developed for the low-density limit in which the smaller clusters appear, capturing the transition from simple single-particle triangular lattices (STL) to low-occupancy cluster crystals with increasing density. 
Generalizing the results above to these questions would open the way to a deeper understanding of cluster self-assembly and pattern formation in a variety of  phenomena, with interesting possible applications in, e.g.,  soft and hard  condensed matter physics.~\cite{dm19sm} 
     
In this work we develop a simple method to predict the emergence of the first cluster phases at low densities for  arbitrary potential in 2D. 
We investigate cluster formation by comparing the energy of the cluster-crystalline configuration with smallest occupancy (i.e. two particles per cluster) with respect to the STL as a function of particle density, at zero temperature.   
This regime in which the first cluster-crystal emerges is what we understand here by "small density" and the precise regime of parameters for which it occurs depends on the specific potential.
We derive a general sufficient criterion for cluster formation by analyzing the interaction energy in terms of sums of lattice components of the Fourier transform of the interactions. 
We confirm the efficacy of this criterion by predicting cluster formation in several examples for bounded and unbounded potentials of direct relevance to different communities.

\section{Method}

\begin{figure}
	\includegraphics[width=1\columnwidth]{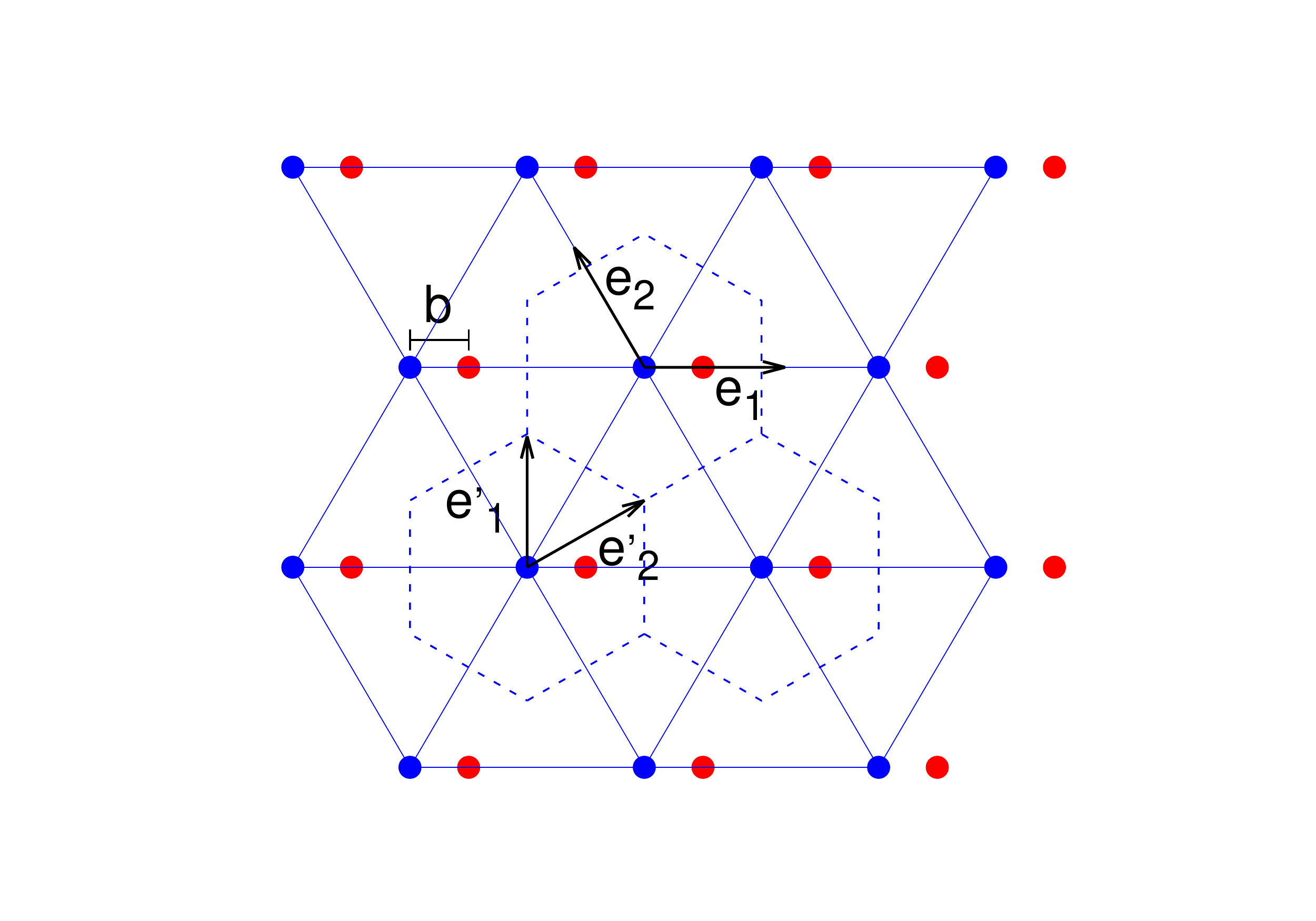}\\
	\caption{
		Scheme of a single-particle triangular lattice (blue circles) where the generating vectors are named $\vec{e}_1$ and $\vec{e}_2$ according to the text.
		The two-particles cluster-crystal is also represented by considering both blue and red circles.
		Following the procedure described in the text, clusters are created by adding a second particle at a distance $b$ in the direction $\vec{e}_1$. 
		}
	\label{tri}
\end{figure}

In two dimensions and zero temperature the low-density limit of the monodisperse crystals corresponds to a triangular array of single particles STL, with a lattice parameter $a_1$ that increases as density is lowered.
The lattice parameter $a_1$ is related to the density (particles per unit area) as $\rho=1/\sqrt{3}a_1^2$.  
This STL is represented in  Fig.~\ref{tri} with color blue, and the positions of the particles can be expressed as 
\begin{equation}
\vec{r}_{nm}=(n\vec{e}_1+m\vec{e}_2)\ a
\label{rmn}
\end{equation}
where $\vec{e}_1=(1,0)$, $\vec{e}_2=(-\frac{1}{2},\frac{\sqrt{3}}{2})$, $n$ and $m$ are integers, and $a=a_1$. 
The energy per particle corresponding to such configuration can be written as
\begin{equation}
\frac{E_1}{N}=\frac{1}{2}\sum_{(n,m)\neq(0,0)}U(r_{nm})
\label{E1}
\end{equation}
where $U(r_{nm})$ is a pair-wise isotropic interaction potential, and  $r_{nm}\equiv |\vec{r}_{nm}|$.

It is convenient to rewrite Eq.~(\ref{E1}) in terms of the Fourier transform of the potential $U(r)$.
However, this Fourier transform is not defined in the general case since many widely used potentials have a strong divergence at $r=0$.
To overcome this issue maintaining the exactness of the analysis the potential $U(r)$ is replaced by   
\begin{equation}
V(r,\xi)=
\left\{
\begin{array}{ll}
U(\xi) & r\leq \xi \\
U(r) & r\geq \xi \\
\end{array} 
\right.,
\label{Vr}
\end{equation}
where $\xi$ represents a short-distances cutoff.
It should be noticed that the sum in Eq.~(\ref{E1}) is independent of the value of $\xi$ as long as $\xi<a_1$.
By construction, the potential in Eq.~(\ref{Vr}) is bounded and consequently its Fourier transform is well defined.
In 2D this transform can be written as 
\begin{equation}
\hat{V}(q,\xi)=2\pi\int_0^\infty dr\ r J_0(qr) V(r,\xi) 
\end{equation}
where $J_0(x)$ is the Bessel function of order zero.

In terms of the regularized potential presented in Eq.~(\ref{Vr}), the energy per particle of our triangular lattice can be rewritten as
\begin{equation}
\frac{E_1}{N}=\frac{1}{2}\left(\sum_{n,m}V(r_{nm},\xi)-V(0,\xi)\right),
\label{E11}
\end{equation}
where the self interaction contribution $n=m=0$ in the sum over the triangular lattice has been explicitly subtracted. 
This expression is thus equivalent to Eq.~(\ref{E1}).

Now we use an identity whose demonstration, for a more general case, can be found in Appendix~\ref{ap1}. 
It relates the sum of an arbitrary function in real space $g(r)$ over a triangular lattice shifted by an arbitrary vector $\vec{\Delta}$, to the Fourier transform $\hat{g}(q)$ of the function $g(r)$,  
\begin{equation}
\sum_{nm} g([n\vec{e}_1+m\vec{e}_2]a+\vec{\Delta})=
\frac{2}{\sqrt{3}a^2}\sum_{nm}\hat{g}(|\vec{q}_{nm}|)\cos(\vec{q}_{nm}\cdot\vec{\Delta}).
\label{eqP}
\end{equation}  
In the right-hand side of this equation the wave vector lattice $\vec{q}_{nm}$ has the form
\begin{equation}
\vec{q}_{nm}=\frac{4\pi}{\sqrt{3}a}(n\pvec{e}'_1+m\pvec{e}'_2),
\label{qnm}
\end{equation} 
where $n$ and $m$ are integer indexes while the lattice vectors are chosen as $\pvec{e}'_1= (0,1)$ and  
$\pvec{e}'_2=(\frac{\sqrt{3}}{2}, \frac{1}{2})$ (see Fig.~\ref{tri}). 

Using Eq.~(\ref{eqP}), the unconstrained sum over particle sites in Eq.~(\ref{E11}) can be rewritten  in terms of the Fourier transform of the regularized potential $V(r,\xi)$
\begin{equation}
\frac{E_1}{N}=\frac{1}{2}\left(\frac{2}{\sqrt{3}a_1^2}\sum_{nm}\hat{V}(|\vec{q}_{nm}|,\xi)-V(0,\xi)\right).
\label{E1q}
\end{equation}
It should be remarked that, while it seems to depend strongly on the parameter $\xi$, Eq.~(\ref{E1q}) is equivalent to Eq.~(\ref{E1}) for any $\xi$ smaller than $a_1$ at $T=0$. 
In this way, both expressions can be used to calculate the energy of the STL at a given density.\\

For systems with CFI, the other relevant configuration at zero temperature and low density is the two-particles cluster-crystal (2CC), i.e. a cluster-crystal with two particles per site.
We will consider that such a configuration is composed by clusters in which particles are separated by a distance $b$ in the direction $\pvec{e}_1$ (see Fig.~\ref{tri}).
This configuration can be visualized as two shifted triangular lattices shown in blue and red in Fig.~\ref{tri}.
The election of the direction $\vec{e}_1$ to place the second particle in the two-particle cluster ansatz, is based on the observation of numerical simulations of the annealing of cluster-forming systems at low temperature.
In general, for any arbitrary potential, the 2CC with clusters along 
one of the main directions of the lattice is not always more stable than that in which clusters orient forming an angle with one of these directions.    
However, as discussed in Appendix~\ref{ap2}, we have tested that this choice does not affect the validity of the analysis, and from now on we assume that it is enough to consider the stability of the 2CC phase along $\vec{e}_1$.

The 2CC have a well defined lattice parameter $a_2$ that depends on the density in the form $\rho=4/\sqrt{3}a_2^2$. This means that at a given density the ratio between the lattice spacing of the 2CC lattice and the STL is given by $a_2/a_1=\sqrt{2}$.      		 
Furthermore, in order to have clusters we consider that the particle separation within a cluster must be less than certain maximum distance $b_\mathrm{max}$ equivalent to half the lattice spacing $b_\mathrm{max}=a_2/2$. 
The value of $b_\mathrm{max}$ will thus 
depend on the density via the lattice parameter $a_2$ of the 2CC.

Following a procedure analogous to that with the STL, we obtain that the energy per particle for the 2CC is given by
\begin{equation}
\frac{E_2}{N}=\frac{1}{2} \sum_{n\neq0,m\neq0} V(|\vec{r}_{nm}|,b)+  \frac{1}{2} \sum_{nm} V(|\vec{r}_{nm} + \vec{b}|,b),
\label{E2q}
\end{equation}
where $\vec{b}=b\pvec{e}_1$ and the vectors $\vec{r}_{nm}$ correspond to Eq.~(\ref{rmn}) with a lattice spacing $a=a_2$. 
It should be noticed that in this case the natural choice for the short distance cutoff is $\xi=b$, as $b$ is the smallest separation between particles in the system. 
Using the identity (\ref{eqP}) the energy per particle can be rewritten as  
\begin{eqnarray}
\nonumber
\frac{E_2}{N}&=&\frac{1}{2}\left(\frac{2}{\sqrt{3}a_2^2}\sum_{nm}\hat{V}(|\vec{q}_{nm}|,b)\right.\\
&+&\left.\frac{2}{\sqrt{3}a_2^2}\sum_{nm}\hat{V}(|\vec{q}_{nm}|,b)\cos(\vec{q}_{nm}\cdot\vec{b})-V(0,b)\right),
\label{E3q}
\end{eqnarray}
with $\vec{q}_{nm}$ evaluated in the lattice parameter $a=a_2$ via Eq.~(\ref{qnm}).

From the definition of $b_\mathrm{max}$, at a fixed density $\rho$ the parameter $b$ can take any value in the interval  $[0,1/\sqrt{\sqrt{3}\rho})$.
With this constraint, the value of $b$ that minimizes the energy of Eq.~(\ref{E3q}), for a given interaction potential, will correspond to the most stable 2CC configuration.

In general, there are two possible behaviors when this energy is minimized in terms of $b$: the minimum can be reached for $b=0$ or for some nontrivial positive value in the interval $[0,1/\sqrt{\sqrt{3}\rho})$. 
Systems interacting via a bounded potential whose Fourier transform have a negative minimum at some nonzero wavevector typically minimize Eq.~(\ref{E3q}) at $b=0$.~\cite{likos01b} 
On the other hand, systems with a strong enough short-distance repulsion normally have its minimum at a positive value of the parameter $b$. 

Direct minimization of the energy shows that, when the 2CC phase is imposed, the parameter $b$ of the 2CC remains roughly constant as density is increased (see Appendix~\ref{ap3}).
Considering only a 2CC, the lattice parameter $a_2$ decreases with increasing density, so that eventually the rather constant value of $b$ equals its maximum value $b_\mathrm{max}=a_2/2$ which has a more pronounced dependency on the density. 
The state in which $b=b_\mathrm{max}$ is thus a reference state that represents a limiting case for cluster-crystals, although can not be properly classified as a cluster state.
Since in this state the value of $b$ is known, a direct analytical calculation of the energy can be made.
In the following we use this state with $b=b_{max}$ to derive an explicit sufficient condition for the transition to occur from single-particle crystal to the two-particle cluster crystal.\\

It is natural to expect that, if a system has the ability to form clusters with increasing density, the 2CC will { usually } be the first cluster-crystal  configuration to emerge. 
A sufficient condition for this cluster crystal to occur can be derived by comparing the energies of the (low density) STL and the corresponding limit case of 2CC with $b=b_\mathrm{max}$.
If the energies of both phases are equal, the following relation has to be fulfilled 
\begin{equation}
 f_1(k,b_\mathrm{max})=\frac{1}{2}\left[f_1\left(\frac{k}{\sqrt{2}},b_\mathrm{max}\right)+f_2\left(\frac{k}{\sqrt{2}},b_\mathrm{max}\right)\right]
 \label{Eq4}
\end{equation}
where
\begin{eqnarray}
\nonumber
 f_1(k)&=&\sum_{mn}\hat{V}(k\sqrt{m^2+n^2+mn},b)-\frac{8\pi^2}{\sqrt{3}k^2}V\left(0,b\right)\\ \nonumber
f_2(k,b)&=&\sum_{mn}\hat{V}(k\sqrt{m^2+n^2+mn},b)\cos(kb\sqrt{3}(n+m)/2),\\
\label{Eq4a}
\end{eqnarray}
with $b_\mathrm{max}=\frac{2\sqrt{2}\pi}{\sqrt{3}k}$
and $k=\frac{4\pi}{\sqrt{3}a_1}$.
Here the value of $k$ depends on the lattice spacing $a_1$ of the STL, which is completely determined by the density $\rho$. 
Additionally, the fact that the functions in the right hand side (RHS) of Eq.~(\ref{Eq4a}) are evaluated in $k/\sqrt{2}$ is a consequence of the relation $a_2=\sqrt{2}a_1$.

The RHS of Eq.~(\ref{Eq4}) is proportional to the energy of the 2CC evaluated for the corresponding $b_\mathrm{max}$. 
The left hand side (LHS) is equally proportional to the energy of the STL, and it does not depend on $b$ since $b_\mathrm{max}=\frac{\sqrt{2}}{2}a_1<a_1$.   
In this way, in order to check if a given potential exhibits a transition to a cluster-crystal phase, we can simply look at the values of the RHS and the LHS of Eq.~(\ref{Eq4}).
If such transition indeed takes place, there have to exist some values of $k$ for which the RHS 
is smaller than the LHS.\\

As said above, the configuration with $b=b_\mathrm{max}$ has to be explicitly excluded from the criterion, as it does not correspond to our definition of cluster-crystal, valid for $b<b_\mathrm{max}$.   
Consequently we additionally monitor the sign of the second derivative $D_2f(k)={\partial^2_bf_2\left(\frac{k}{\sqrt{2}},b\right)}\vert_{b=b_{\mathrm{max}}}$. 
This guarantees that the configuration with $b=b_{max}$ is unstable with respect to the formation of cluster-crystals.

For actual cluster-forming potentials it has to exist a density (or $k$) value, for which the RHS is smaller than the LHS in Eq.~(\ref{Eq4}), and $D_2f(k)<0$.
These two relations forms a sufficient condition for the determination of the cluster-forming ability of any potential.

\section{Results}
The sufficient criterion is now tested in a number of well-known potentials from different contexts in the literature.
Some of them have the ability to form clusters while others remain in the triangular array configuration for all densities.
We discuss separately the CFI and non-CFI examples. 

\subsection{Cluster Forming Interactions}
\label{CF}

\begin{figure}[H]
	\includegraphics[width=0.95\columnwidth]{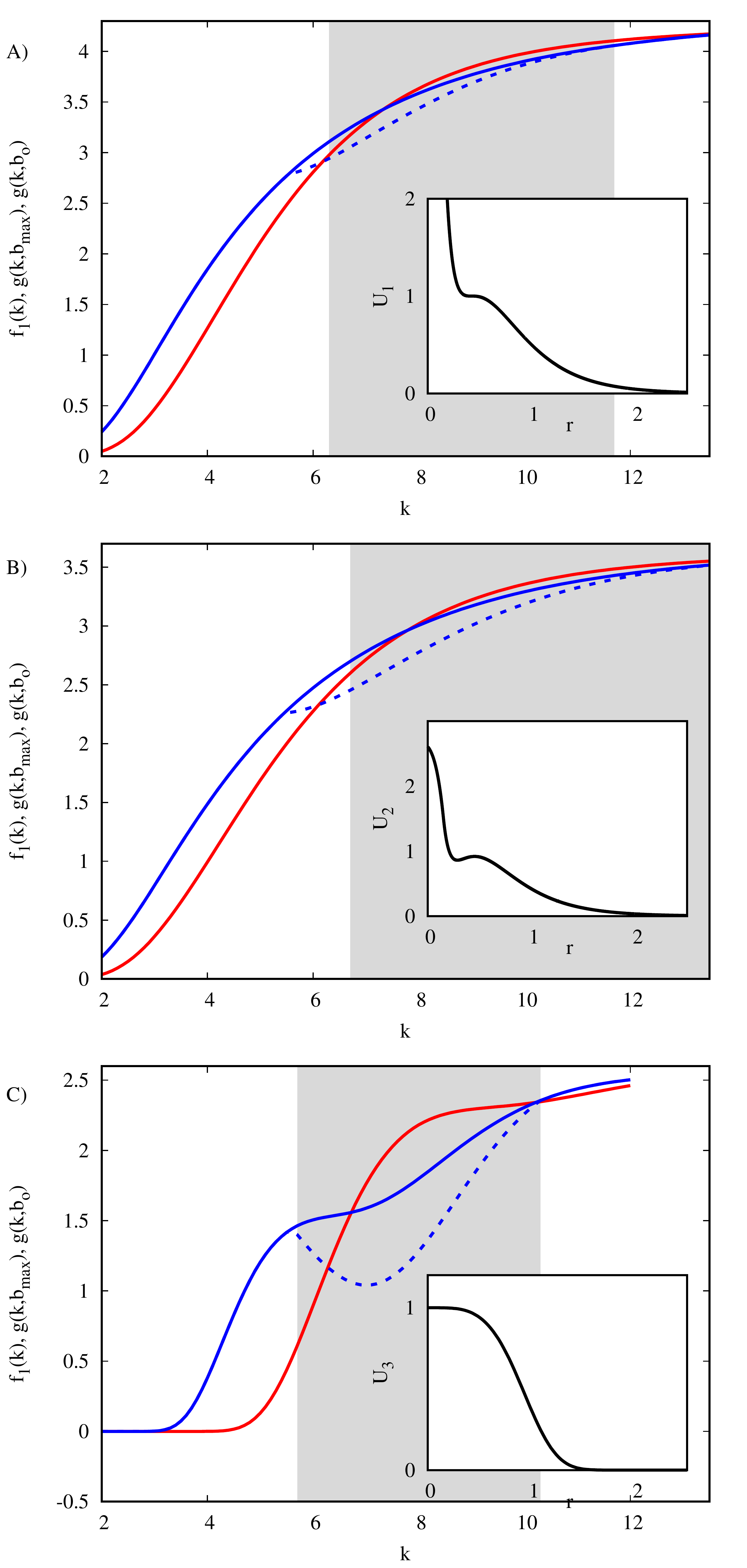}
	\caption{
		Functions $f_1(k)$ (solid red curve), $g(k,b_\mathrm{max})$  (solid blue curve) and $g(k,b_\mathrm{o})$  (dashed blue curve) plotted as a function of the wave vector $k$, for three examples of CFI.
		A) Potential $U_1$ from Eq.~(\ref{v1}), B) Potential $U_2$ from Eq.~(\ref{v2}) and C) Potential $U_3$ from Eq.~(\ref{v3}).      
		Shaded regions correspond to the values of $k$ in which $D_2f(k)<0$. As explained in the text, the existence of extended $k$ (or density) regions in which simultaneously $g(k,b_\mathrm{max})<f_1(k,b_\mathrm{max})$ and $D_2f(k)<0$ implies the stability of the cluster phase, which then can be used as a sufficient criterion to identify cluster forming pair-wise potentials. In all panels the insets are plots of the potential corresponding to each main panel.    
		}
	\label{fig2}
\end{figure}

{ The first example potential $U_1$ describes the asymptotically-derived  
long-range  interaction between vortices in ``type-1.5'' superconductors, and layered systems which are multiband superconductors that have multiple coherence lengths some larger and some smaller than magnetic field penetration length.~\cite{bs11,johan2,varney2013hierarchical,meng2016phase,dm17,silaev1,moshchalkov,Ray.Gibbs.ea:14} 
 In the simplest case the
system has only one repulsive length scale, but multiple repulsive length scales arise in layered type-1.5  systems as well as in anisotropic multiband superconductors \cite{winyard2019hierarchies}.}
It is represented in the inset of Fig.~\ref{fig2}A, and its mathematical form is given by

\begin{equation}
	U_1(r)=\sum_{i=1}^2 B_i^2 K_0(r/\lambda_i) - C_i^2 K_0(r/\xi_i) 
	\label{v1}
\end{equation}
where $K_0(r)$ is the modified Bessel function of the second kind, 
$B_1=3.05386$, $\lambda_1=0.420893$, $C_1=6.27334$, $\xi_1=0.210447$, $B_2=6.6922$, $\lambda_2=0.140296$ and  $C_2=0$.
This set of parameters is used in Ref.~[\cite{dm17}] and corresponds to an unbounded potential for which the Fourier transform is not defined. 
Consequently, the well-known criterion developed by Likos et~al.~\cite{likos01} is not applicable.

The second example potential $U_2$ has the form
\begin{equation}
U_2(r)=
\left\{
\begin{array}{ll}
A - D \ (0.15+r)^4  & r\leq 0.15 \\
\sum_{i=1}^2 B_i^2 K_0(r/\lambda_i) - C_i^2 K_0(r/\xi_i) & r\geq 0.15 \\
\end{array} 
\right.
\label{v2}
\end{equation}
with $A=2.67956$, $D=144.376$, $B_1=2.7$, $\lambda_1=0.42$, $C_1=6.0159$, $\xi_1=0.196$, $B_2= 6.1$, $\lambda_2=0.1405$ and  $C_2=0$. 
This interaction, represented in the inset of Fig.~\ref{fig2}B, is similar to $U_1$ and considered in the same context.\cite{dm19w2}
The main difference is that, for $U_2$, the potential energy is bounded at short distances, which means that the Fourier transform of the potential is well defined. 
Nonetheless, since it does not exhibit any negative minimum, this is a case of CFI that goes beyond the Likos' criterion. 
The latter ensures that the potential $U_2$ does not develop cluster configurations at sufficiently high density, which is correct.
However, there is a rich cluster-crystal phenomenology
accessible via computer simulations at moderate densities that is not predicted.~\cite{dm19w2}         
In this way, potentials $U_1$ and $U_2$ are representative of a wide class of systems for which there is no criterion to predict the cluster-forming character.

We additionally include in the analysis the generalized exponential model of index 4 (GEM4),\cite{likos08,prestipino14} defined by an interaction potential of the form
\begin{equation}
 U_3(r)=\exp(-r^4),
 \label{v3}
\end{equation}
which is represented in the inset of Fig.~\ref{fig2}C.
This potential has been widely used in the context of colloidal and polymeric systems.\cite{coslovich11,mladek08}   
This is a bounded potential having a Fourier transform  with a negative minimum at some wave vector. 
In this case the Likos' criterion is applicable and the cluster-forming ability of the potential is known a priori.
Nevertheless, the condition of Eq.~(\ref{Eq4}) is applicable as well, and able to detect the transition from the STL to the 2CC (see Appendix~\ref{ap31}).\\

In order to simplify the notation we define the function $g(k,b)$ as the RHS of Eq.~(\ref{Eq4})
\begin{equation}
 g(k,b)=\frac{1}{2}\left[f_1\left(\frac{k}{\sqrt{2}},b\right)+f_2\left(\frac{k}{\sqrt{2}},b\right)\right],
\end{equation}
and calculate the curves $g(k,b_\mathrm{max})$ and $g(k,b_\mathrm{o})$. As mentioned before, $b_\mathrm{max}=\frac{2\sqrt{2}\pi}{\sqrt{3}k}$, while $b_\mathrm{o}$ is a function of $k$, such that $g(k,b_\mathrm{o})$ corresponds to the minimum value of $g(k,b)$, taking $b$ as a variational parameter in the interval $[0,b_\mathrm{max}]$; that is to say $g(k,b_\mathrm{o})=\mathrm{Min}_b[g(k,b)]$.

In Fig.~\ref{fig2} both the LHS of Eq.~(\ref{Eq4}) [$f_1(k)$, solid red curve] and the RHS  [$g(k,b_\mathrm{max})$, solid blue curve] are plotted as a function of the wave vector $k$ for each example potential $U(r)$ given above.
For comparison we also plot the function $g(k,b_\mathrm{o})$ (dashed blue line), which, as said above, is calculated for the value of $b$ that mimimizes $g(k,b)$ in the RHS of Eq.~(\ref{Eq4}). 
Additionally, in all panels, the shaded region corresponds to the values of $k$ in which $D_2f(k)<0$. 

For the repulsive potentials of interest in this work, the energy of the system increases with the increase of the density. Increasing density is equivalent to  diminish the lattice spacing $a_1$ or to increase the value of $k$.
As a consequence, the functions $f_1(k)$ and $g(k,b_\mathrm{max})$ grow with the value of $k$ as shown in Fig.~\ref{fig2}.
In all panels, for small $k$ values the red curves are always lower than the blue curves, reflecting the fact that, at low densities, the energy of the STL is smaller than that of the 2CC. In all cases, however, the ground state changes from the STL to a 2CC at certain values of density.

This change is estimated in the figure by the intersection of the red and solid blue curves, corresponding to the value of $k$ for which Eq.~(\ref{Eq4}) is fulfilled. 
The values of the wave vectors $k$ at these points define the lattice spacing and density at which the energy of the STL equals the energy of a 2CC formed by particles separated a distance $b_\mathrm{max}$ within the clusters.

In turn, the points where the solid red and the dashed blue curves cross define the $k$ value at which the energy of the STL equals that of the 2CC with minimal energy, i.e. the 2CC energy obtained from Eq.~\eqref{E3q} by setting $b=b_\mathrm{o}$. 
As can be seen from the figure, this crossing occurs for values of $k$ smaller than the crossing described in the above paragraph between the STL line (solid red) with the 2CC with cluster separation $b=b_\mathrm{max}$ (solid blue).
However, the difference between the blue curves $g(k,b_\mathrm{max})$ and $g(k,b_\mathrm{o})$ decreases by further increasing $k$. 
This is a confirmation of the above mentioned tendency to maintain an almost fixed separation between particles inside a cluster . \\

The verification that an extended region of wave vectors $k$ exists in which $g(k,b_\mathrm{o})<f_1(k,b_\mathrm{max})$ proves the existence of clusters states for the potentials under analysis. Our work provides a sufficient criterion to ensure the existence of cluster states without performing such calculations: it is enough to verify the existence of $k$ regions in which simultaneously the conditions $f_1(k,b_\mathrm{max})>g(k,b_\mathrm{max})$ and $D_2f(k)<0$ are fulfilled. As shown in Fig.~\ref{fig2}, the developed criterion correctly predicts the cluster forming character of the three interaction potentials considered here, supporting our original statement about its generality.

\begin{figure}
	\includegraphics[width=0.95\columnwidth]{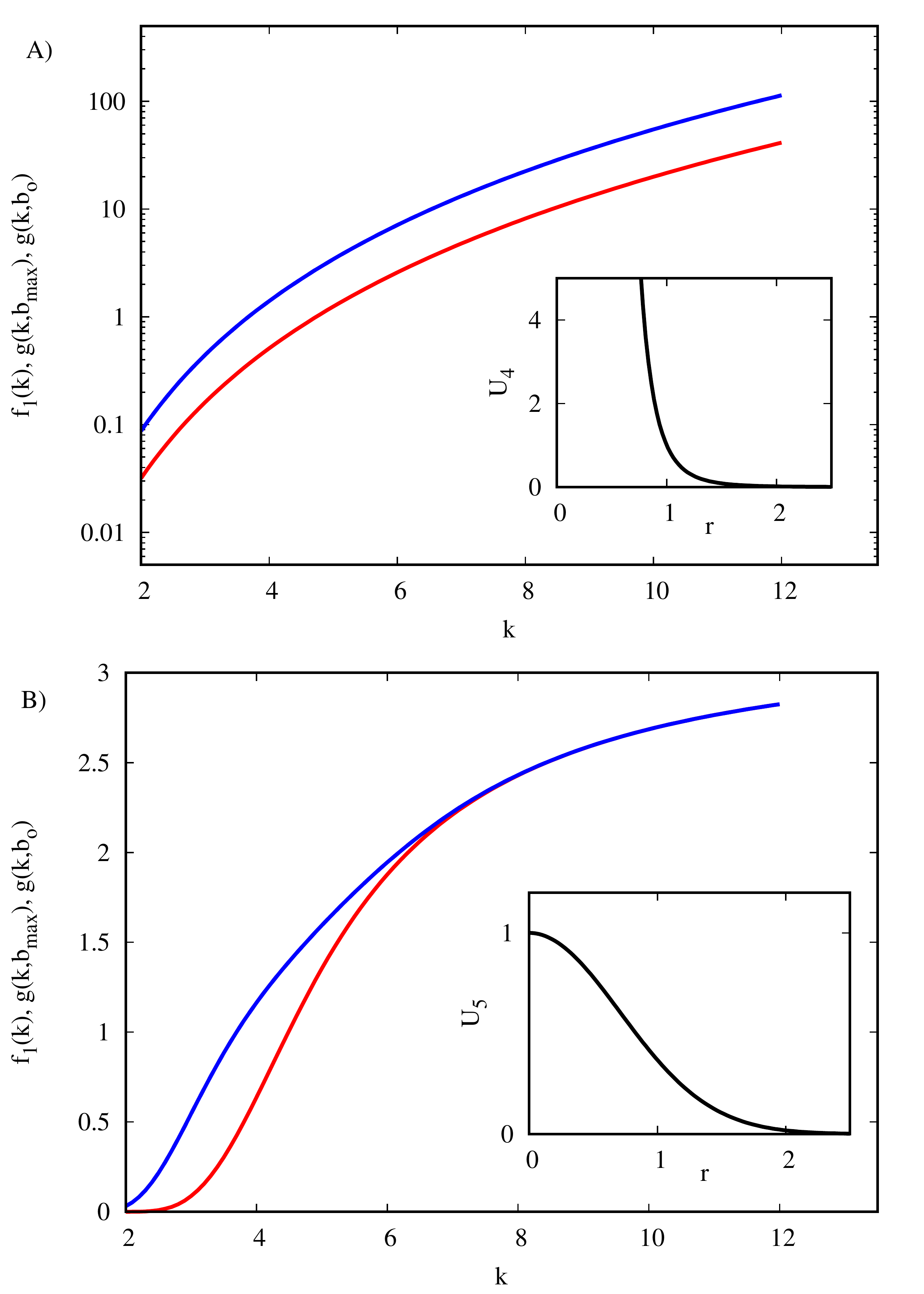}
	\caption{
		Energetic terms $f_1(k)$ (red curve), $g(k,b_\mathrm{max})$  (solid blue curve) and $g(k,b_\mathrm{o})$  (dashed blue curve) for three examples of non-CFI.
		A) Potential $U_4$ from Eq.~(\ref{v4}) and B) Potential $U_5$ from Eq.~(\ref{v5}).      
		The insets are plots of the corresponding potential. As can be observed in the figures none of the conditions forming the developed criterion for identifying cluster forming potentials is fulfilled by the considered potentials. 
		}
	\label{fig3}	
\end{figure}

\subsection{Non Cluster Forming Interactions}

The test of the criterion of Eq.~(\ref{Eq4}) is completed with the inclusion of two non-CFI of bounded and unbounded nature.
As a model of unbounded potential we consider the Van der Waals interaction   
\begin{equation}
U_4(r)=\frac{1}{r^6}.
\label{v4}
\end{equation}
This potential is used in the context of molecular and atomic interactions\cite{cinti14} and is represented in the inset of Fig.~\ref{fig3}A.

We also look at the so called Gaussian Core Model (GCM) defined by the potential
\begin{equation}
 U_5(r)=\exp(-r^2).
 \label{v5}
\end{equation}
This interaction is commonly used to study colloidal systems\cite{prestipino05} and is represented in the inset of Fig.~\ref{fig3}B. 
It is a bounded potential for which the Likos' criterion predicts no cluster-crystal phases at high density. 

Potentials $U_4$ and $U_5$ do not develop cluster phases for any density, as it is known from many numerical studies.
Therefore, in both cases, the curves of $g(k,b_\mathrm{max})$ and $g(k,b_\mathrm{o})$ superposes in Fig.~\ref{fig3}.
Since the potentials are non-CFI, the system always rearranges in states in which particles remain as far as possible from each other.
In this situation the configuration that minimizes the energy is the triangular lattice, and consequently the curves of $f_1(k)$ are the lowest for all values of $k$.

Interestingly, in Fig.~\ref{fig3}B, the difference between the curves of $f_1(k)$ and $g(k,b_\mathrm{max})$ is asymptotically zero for large values of $k$, i.e. high density. 
This is a consequence of the particular properties of the family of GEM$\alpha$ potentials. 
For values of $\alpha$ in the region $\alpha\in(2,\infty)$ the Fourier transform of the GEM$\alpha$ potential is not positive definite, i.e. the potential is cluster-forming.
In the GEM2 case,  also known as GCM, this value of $\alpha=2$ corresponds exactly to the boundary of the cluster-forming ability.
It means that for any $\alpha=2+\epsilon$ with $\epsilon>0$, the corresponding GEM$\alpha$ potential will be a CFI.
Therefore, for any $\epsilon>0$ the intersection point of $f_1(k)$ and $g(k,b_\mathrm{max})$ have to occur at a finite value of $k$.

The potential of the GCM, and its energy curves shown in Fig.~\ref{fig3}B, are thus a limiting case for which this point is located at an infinite wave vector. 
As $\epsilon\rightarrow0$, the position $k_m$ of the negative minimum of the Fourier transform of the potential diverges $k_m(\epsilon)\rightarrow \infty$; while the minimum of the potential $\hat{V}(k_m(\epsilon))$ approaches zero from below. 
Previous works have established that, at least within the mean-field approximation, the transition from the STL to the 2CC state scales as $k_m^d$, being $d$ the dimension of the system.~\cite{likos08} 
This argument clarifies the above discussion about the behavior of $f_1(k)$ and $g(k,b_\mathrm{max})$ in the limit of $\epsilon\rightarrow0$.

\section{Concluding remarks}

A sufficient criterion for the cluster-forming ability of arbitrary interactions in 2D is derived in terms of the Fourier transform of a regularized version of the potential.
The basis of the criterion is a zero-temperature comparison of the energy unbalance between the single-particle lattice and the first cluster-crystal configuration at small density. 
As a result, a sufficient condition for the emergence of cluster phases is obtained through a sum of lattice components that has to be evaluated at a fixed density.

The criterion allows to predict the cluster-forming ability of all kind of pair-wise interactions.
This advantage complements the well-known Likos's criterion for bounded potentials whose  predictions apply to the regime of high density.   
In this way, systems with unbounded interactions or behaving as cluster-crystals only at moderate densities are the primary context of applicability of this work.
{ One of the applications is for the structure formation of vortex matter in type-1.5 superconductors. For those system   the asymptotically-derived intervortex interaction potentials allow to treat
a dilute system of vortices as particles that have multi-scale forces at long-range. The long-range structure  of asymptotic interaction potentials is consistent with microscopic theory and dictates structure formation in low magnetic fields, but the potential have a divergence at the origin.}

The accuracy of the predictions regarding the cluster-forming ability was tested with several potential interactions used in the literature for which the behavior is known from computer simulations.
In all cases the criterion correctly predicted the cluster formation, allowing to additionally obtain a reasonable estimation of the density at which the first cluster-crystal appears.

\appendix

\section{Lattice sum identity}
\label{ap1}
In order to prove the identity of Eq.~(\ref{eqP}) let us consider the sum
\begin{equation}
 S=\sum_{n,m}g\Big((n\vec{e_1}+m\vec{e}_2)a+\vec{\Delta}\Big),
\end{equation}
where $\vec{e}_1=(1,0)$ and $\vec{e}_2=(\cos{\theta},\sin\theta)$, represent the basis vectors of the lattice over witch the sum is performed. The vector $\vec{\Delta}$ represent an arbitrary shift of the lattice over which the summation is performed. 

The goal is to obtain an equivalent expression for $S$ in terms of the Fourier transform of $g(\vec{r})$ assuming that such function exists.
Note that in our case this is always possible since $g(\vec{r})$ remains finite in all points of the summation lattice.

Considering the following definition of Fourier transforms
\begin{eqnarray}
 \hat{g}(\vec{k})&=&\int d^2r\ e^{-i\vec{k}\cdot\vec{r}}g(\vec{r})\\
 g(\vec{r})&=&\int\frac{d^2k}{(2\pi)^2}\ e^{i\vec{k}\cdot\vec{r}}\hat{g}(\vec{k}),
\end{eqnarray}
the original sum can be rewritten as
\begin{equation}
 S=\int\frac{d^2k}{(2\pi)^2}\sum_{n,m} e^{i\vec{k}\cdot((n\vec{e_1}+m\vec{e}_2)a+\vec{\Delta})}\hat{g}(\vec{k}).
 \label{fsum}
\end{equation}
And taking advantage of the Dirac comb identity 
\begin{equation}
\sum_n e^{inx}=(2\pi)\sum_n\delta(x-2\pi n),
\end{equation}
the summation in Eq.~(\ref{fsum}) can be performed to reach to the following expression for $S$
\begin{eqnarray}
\nonumber
 S&=&\int\frac{d^2k}{(2\pi)^2}e^{i\vec{k}\cdot\vec{\Delta}} \hat{g}(\vec{k})\\
 &\times&(2\pi)^2\sum_{n,m} \delta(\vec{k}\cdot\vec{e}_1a-2\pi n)\delta(\vec{k}\cdot\vec{e}_2a-2\pi m).
 \label{fsum1}
\end{eqnarray}
Now it can be directly integrated over momenta yielding
\begin{equation}
 S=\frac{1}{a^2\vert\sin(\theta)\vert}\sum_{n,m}\hat{g}(\vec{k}_{n,m})\cos(\vec{k}_{n,m}\cdot\vec{\Delta}),
\end{equation}
where
\begin{equation}
\vec{k}_{n,m}=\frac{2\pi}{a\vert\sin(\theta)\vert}(n\pvec{e}'_1+n\pvec{e}'_2),
\end{equation}
with  $\pvec{e}'_1=(0,1)$ and  $\pvec{e}'_2=(\sin(\theta),-\cos(\theta))$. 

In this last step we have also considered that, since $g(\vec{r})$ is a real function, the complex exponential in Eq.~(\ref{fsum1}) can be replaced by its real part. 
It is now straight forward to check the validity of Eq.~(\ref{eqP}) just by considering that our sum is performed over a triangular lattice for which $\theta=2\pi/3$.

\section{Cluster's orientation in the 2CC}
\label{ap2}

For certain interactions, the orientation of the two-particle clusters in the 2CC ground state may vary as density is increased adopting one of two possible cases: clusters oriented along one of the main directions of the lattice (2CC1), or clusters aligned forming an angle of  $\pi/6$ with one of the main directions of the lattice (2CC2). 
As a systematic case study we calculate here the ground-state of a cluster-forming potential frustrated by a tunable repulsion, allowing to vary the relative size of the cluster with respect to the lattice size. 
This type of numerical analysis sheds light on the assumption made in the text of considering only the 2CC1 phase as the cluster-forming predictor.

Let's focus on the following potential interaction
\begin{equation}
V(r)=\exp(-r^4)+Cr^{-6}, 
\label{v}
\end{equation}
where $C$ is a parameter characterizing the relative intensity of the frustrating potential. 
This potential has been studied before in the context of colloids, ~\cite{caprini18} and has the advantage that small values of $C$ can tune the size of the two-particle clusters.
In this way Eq.~(\ref{v}) serves as a toy model to explore the preferential orientation of the 2CC phase when the interplay between the GEM4 cluster-forming term $\exp(-r^4)$ and the frustrating non-cluster-forming Van der Vals repulsion $r^{-6}$ compete.

\begin{figure}[ht]
	\includegraphics[width=0.95\columnwidth]{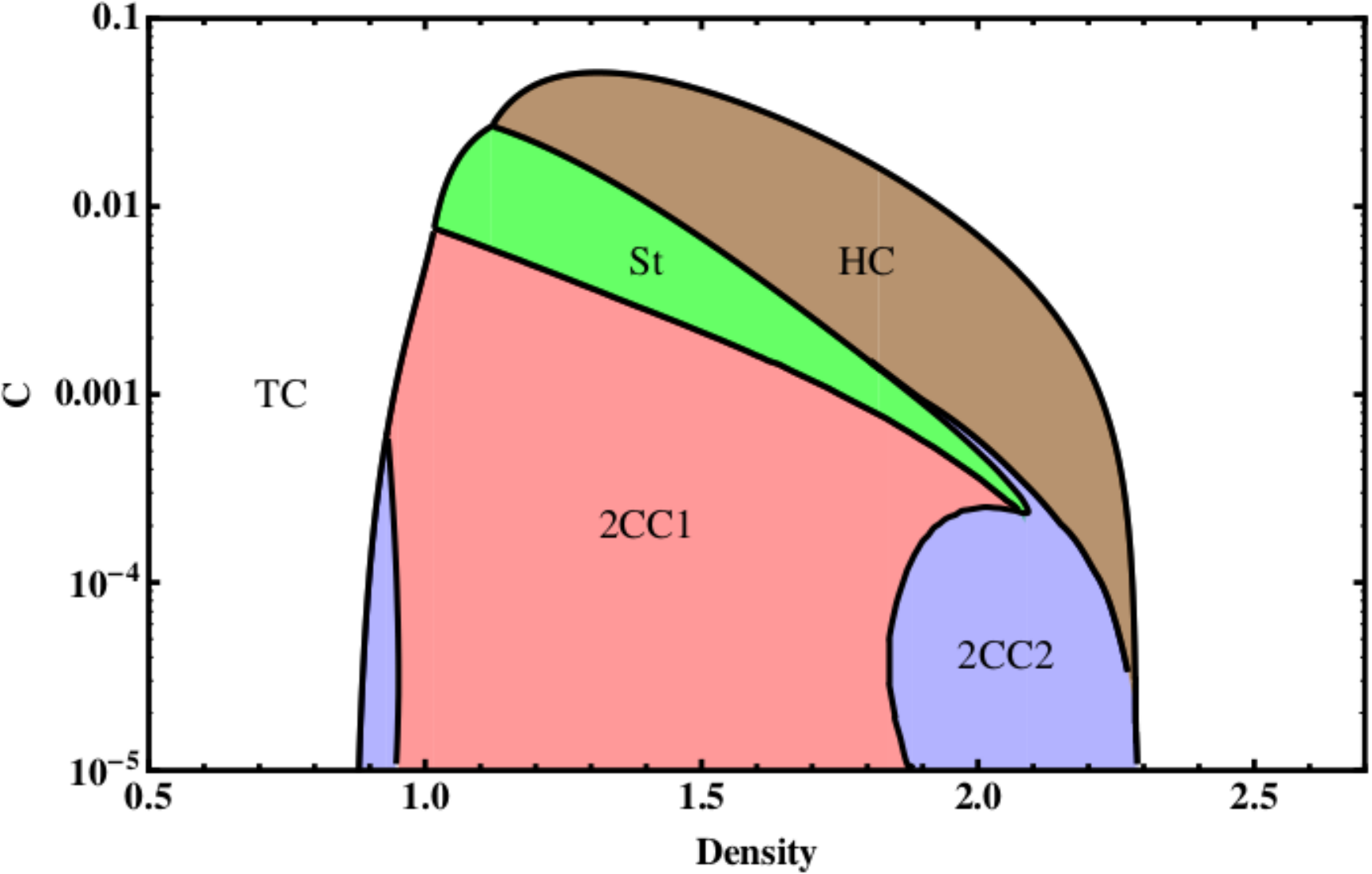}
	\caption{
		Simplified ground-state phase diagram for the potential of Eq.~(\ref{v}). 
		Regions corresponding to 2CC1, 2CC2, stripes (St) and honeycomb (HC) are obtained by energy minimization. 
		The stripes (green) region actually represents a crossover phase between the perfect stripes and the Honeycomb phase. 
		In this crossover the second particle of the cluster moves continuously from $(1/2,0)a_2$, in the boundary with 2CC1,  to $(1/2,1/(2\sqrt{3}))a_2$ in the boundary with the Honeycomb phase.
}
	\label{fig4}	
\end{figure}

A direct energy minimization of the cluster configurations allowing arbitrary orientation demonstrated that only 2CC1 and 2CC2 minimize 
locally  the clusters energy. 
We constructed then a simplified ground-state phase diagram by comparing only the energy of the STL with that of 2CC1 and 2CC2, as well as its limiting cases in which the size of the cluster equals the distance between neighboring clusters in the crystal, these are the stripes and honeycomb phases respectively.
For simplicity we do not determine the possible coexistence regions between different phases.


In Fig.~\ref{fig4} the obtained phase diagram is shown. 
For moderate intensities of the frustrating term ($C\sim 10^{-3}$) the cluster size is comparable to half the lattice spacing.
In this case the transition with increasing density occurs directly from the STL to the 2CC1 phase. 
The repulsion from neighbors of different triangular sub-lattice (see Fig~\ref{tri}) is so strong that each particle benefits from an approximation to its first neighbor, minimizing the repulsion energy with the rest. 

On the other hand, for the regime with low frustration ($C\sim 10^{-4}$)  a transition from the STL to 2CC2 takes place. 
In this regime the cluster sizes are small and the repulsion between neighbors in different sub-lattices is such that the system
minimizes its orientational energy by setting each particle at an equal distance of its first two neighbors in the other sub-lattice. 

More relevant for our discussion, as the density is increased and the cluster size grows, a second transition to a 2CC1 phase takes place. 
This means that, even in the regime of low frustration,
for values of the density large enough we always recover the validity of our assumption about the stability of the 2CC1 phase. 
The 2CC2 phase in this regime of the frustrating potential only exists in a relatively small region of densities and, more important, 
we verified that the energy difference between the 2CC1 and 2CC2 phases is small in comparison with their individual values.  
This means that, although for the purposes of determining an accurate phase diagram it is not correct, for the purpose of establishing the existence of a 2CC phase the 2CC1 ansatz is enough.

\section{Cluster's extension in the 2CC}
\label{ap3}

The distance between particles within the clusters has in general a non-trivial dependency on the density.
For the set of cluster-forming potentials discussed in this work, only $U_1(r)$ and $U_2(r)$ form clusters of finite size, due to a strong enough repulsion acting at small $r$. 
For the potential $U_3(r)$ the zero-temperature clusters are formed by superimposed particles, i.e. the equilibrium distance between the particles forming a cluster is always zero.    

By direct energy minimization of the cluster configurations, allowing at most two particles per cluster, five phases are identified:
STL, 2CC1 (2CC with clusters oriented along $\vec{e}_1$), 2CC2 (2CC with clusters oriented along $\pvec{e}'_2$), stripes (St) and honeycomb (Hc).
Note that St and Hc are the limiting cases of 2CC1 and 2CC2 respectively, when the size of the cluster equals the distance between neighboring clusters in the crystal.

\begin{figure}
\includegraphics[width=0.95\columnwidth]{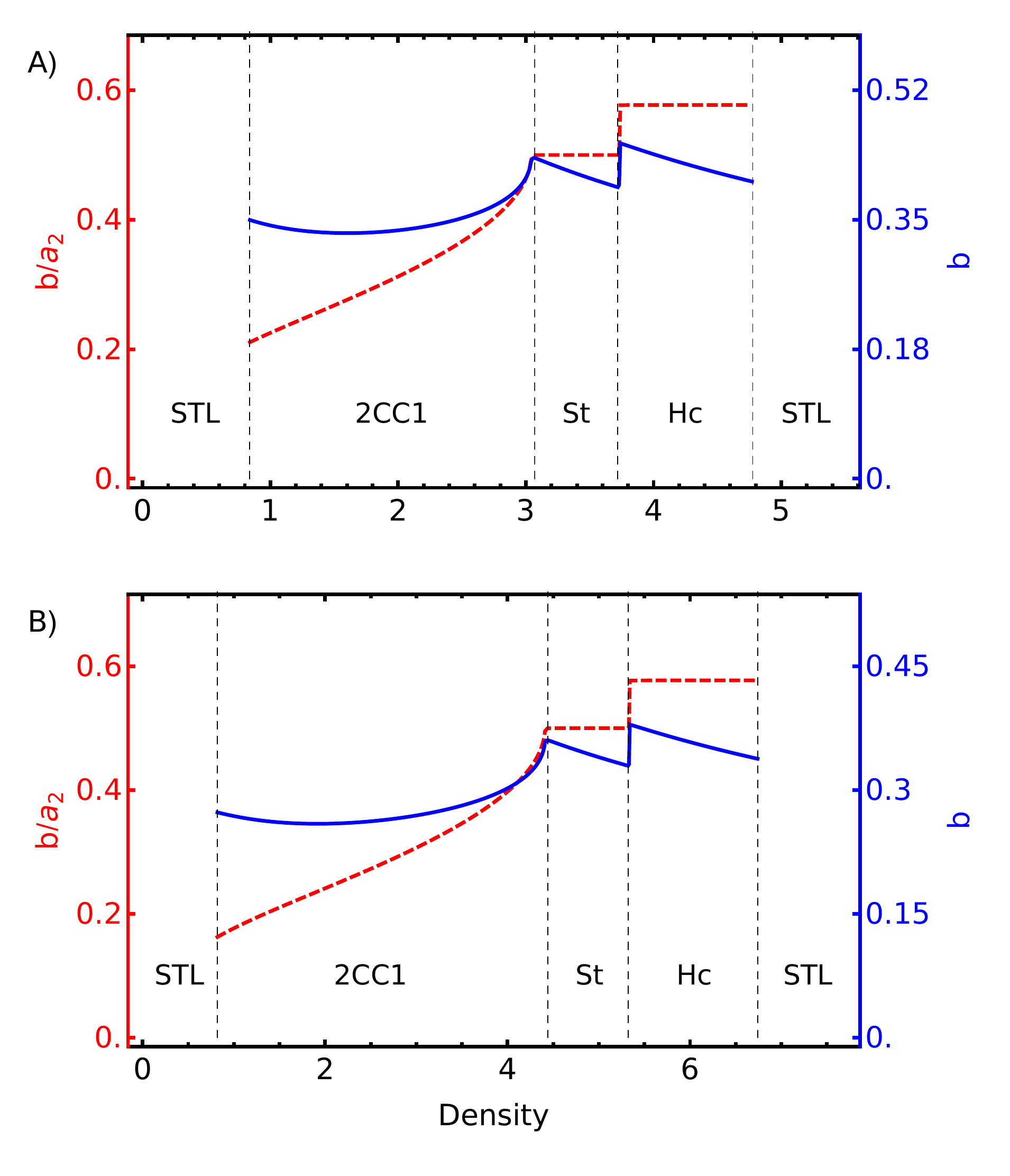}
	\caption{
		Behavior of the equilibrium distance $b_o(\rho)$ between particles within clusters for the potentials $U_1(r)$ (top panel, Fig.~\ref{fig5} A) and $U_2(r)$ (bottom panel, Fig.~\ref{fig5} B). The blue curves and its associated scale presented in the right of the figures, are related with the absolute distance between particles within a cluster. Moreover the red curves represent the distance between particles in units of the 2CC lattice spacing $a_2(\rho)$, the associated scale in this case is presented in red at the left side of the figures. The sequence of phases obtained as the density is increased: STL, 2CC1, St, Hc and STL is presented in the figures and their phase boundaries are represented by dashed vertical lines.   
}
	\label{fig5}	
\end{figure}

In Fig.~\ref{fig5} the behavior of this intra-cluster equilibrium distance $b_o$ is shown for the potentials $U_1(r)$ (Fig.~\ref{fig5} A) and $U_2(r)$ (Fig.~\ref{fig5} B). 
In both panels, $b_o$ is represented by a solid blue line.
In turn, the red dashed line represents the relative value of this equilibrium distance in the units of the lattice spacing $a_2$. 
In both cases we marked the density regions corresponding to the different ground-state phases available under the constraint that only one or two-particle states are allowed.

As can be observed, the equilibrium distance presents a modest non-monotonic variation in the 2CC1 region. 
In this regime, however, the ratio $b_o/a_2$  grows up steadily as the density is increased, until it reaches its maximum possible value $b_o/a_2=1/2$.
The latter corresponds to the limiting case of a 2CC1, which is a stripe state.~\cite{malescio03}
The stripes do not properly represent a cluster configuration and were therefore excluded in our definition of clusters.
The contiguous plateau of $b_o/a_2$  when density is increased are characteristic of the St and Hc phases, as well as the $\rho^{-1/2}$ decay of the absolute value of $b_o$.    

\subsection{Ultrasoft potentials}
\label{ap31}

For potentials that form clusters of superimposed particles at zero temperature, like $U_3(r)$, the quantity $b_o/a_2$ remains zero during the whole 2CC phase.
The criterion presented in this work, however, is still valid for those cases, as it is shown in section \ref{CF}. 

In order to understand the validity of the criterion in these cases let's consider a cluster-forming potential $U_0(r)$, in which clusters are formed by superimposed particles at zero temperature.
It is natural to expect that this cluster-forming character does not disappear if the potential is altered by adding a small-enough hard-core term in the form
\begin{equation}
U_\epsilon(r)=
\left\{
\begin{array}{ll}
\infty~,  & r\leq \epsilon \\

0 \ \  , & r\geq \epsilon. \\
\end{array} 
\right.
\label{vu}
\end{equation}
And now the system with potential 
\begin{equation}
U(r)=U_0(r)+U_\epsilon(r)
\end{equation}
should satisfy our criterion since, by construction, it is a cluster-forming potential of the kind discussed before. 

At the same time, if the criterion is able to detect the cluster formation for the potential $U(r)$ it is because there exists a density $\rho_0$ at which the energy of the triangular lattice $E^U_{STL}(\rho_0)$
equals the energy of the 2CC configuration $E^U_{2CC}(\rho_0)$, that is
\begin{equation}
E^U_{STL}(\rho_0)=E^U_{2CC}(\rho_0).
\end{equation}
The energy of the STL is not affected by the addition of a small-enough hard core, 
\begin{equation}
E^U_{STL}(\rho)=E^{U_0}_{STL}(\rho).
\end{equation}
And, since the superimposed 2CC configuration is the ground state of $U_0$, the inclusion of a hard-core potential makes the energy of the new 2CC ground-state always greater than (or equal to) its superimposed counterpart. 
Therefore 
\begin{equation}
E^{U_0}_{2CC}(\rho_0)\leq E^U_{2CC}(\rho_0).
\end{equation}
and consequently, 
\begin{equation}
E^{U_0}_{2CC}(\rho_0)\leq E^{U_0}_{STL}(\rho_0).
\end{equation}
which proves the validity of the criterion for the potential $U_0(r)$.

\section*{Acknowledgements}

The work was supported by the Swedish Research Council Grants No. 642-2013-7837, 2016-06122, 2018-03659 and G\"{o}ran Gustafsson  Foundation  for  Research  in  Natural  Sciences  and  Medicine
and Olle Engkvists Stiftelse.
A.M.C. acknowledges financial support from Funda\c{c}\~ao de Amparo \`a Pesquisa de Santa Catarina, Brazil (Fapesc).
G.P. acknowledges financial support from icFRC, the University of Strasbourg Institute of Advanced Studies (USIAS) and the Institut Universitaire de France (IUF).

\section*{References}

%

\end{document}